\documentstyle[prl,twocolumn,epsf,floats,aps]{revtex}

\newcommand{\beq}{\begin{equation}}
\newcommand{\eeq}{\end{equation}}
\newcommand{\bea}{\begin{eqnarray}}
\newcommand{\eea}{\end{eqnarray}}
\newcommand{\beano}{\begin{eqnarray*}}
\newcommand{\eeano}{\end{eqnarray*}}

\newcommand{\bb}{\bar b}
\newcommand{\fb}{\bar f}

\newcommand{\fbd}{\fb^{\dagger}}

\newcommand{\fbph}{\fb^{\vphantom{\dagger}}}

\newcommand{\up}{\uparrow}
\newcommand{\down}{\downarrow}

\newcommand{\ra}{\rangle}
\newcommand{\la}{\langle} 

\begin{document}
\draft 
 
\twocolumn[\hsize\textwidth\columnwidth\hsize\csname
@twocolumnfalse\endcsname
 
\title{Exact exponents for the spin quantum Hall transition}
\author{Ilya A. Gruzberg$^1$, Andreas W. W. Ludwig$^{1,2}$, and N. Read$^3$}
\address{$^1$Institute for Theoretical Physics, University of California, 
Santa Barbara, CA 93106-4030 \\
$^2$\cite{permaddr}Physics Department, University of California, Santa Barbara, 
CA 93106-4030 \\
$^3$Department of Physics, Yale University, P.O. Box 208120, New
Haven, CT 06520-8120}
\date{\today}
\maketitle
 
\begin{abstract}
We consider the spin quantum Hall transition which may occur in disordered 
superconductors with unbroken SU$(2)$ spin-rotation symmetry 
but broken time-reversal symmetry. Using supersymmetry, we 
map a model for this transition onto the two-dimensional percolation problem.
The anisotropic limit is an sl$(2|1)$ supersymmetric spin chain.
The mapping gives exact values for critical exponents associated with
disorder-averages of several observables in good agreement with recent 
numerical results. 
\end{abstract}

\pacs{PACS numbers: 73.40.Hm, 73.20.Fz, 72.15.Rn} 
]

Noninteracting electrons with disorder, and the ensuing metal-insulator
transitions, have been studied for several decades, and are usually divided into
just three classes by symmetry considerations.  Recently, the ideas have been
extended to quasiparticles in disordered superconductors, for which the particle
number is not conserved at the mean field level.  Several more symmetry classes
have been found \cite{az}.  One of these, denoted class C in Ref.\ \cite{az}, is
of particular interest \cite{sfbn,sf,khac,brad}.  This is the case in which
time-reversal symmetry is broken but global SU(2) spin-rotation symmetry is not,
and spin transport can be studied.  In two dimensions (2D) it can occur in
$d$-wave superconductors.  Within class C, a delocalization transition is
possible in which the quantized Hall conductivity {\em for spin} changes by two
units, resembling the usual quantum Hall (QH) transition but in a different
universality class.  When a Zeeman term is introduced which breaks the SU(2)
symmetry down to U(1), the transition splits into two that are each in the usual
QH universality class.

In this paper we present {\em exact} results for a recent model \cite{khac,brad}
for the spin QH transition, in class C, in a system of noninteracting
quasiparticles in 2D.  We use a supersymmetry (SUSY) representation of such
models, considered previously \cite{grs}, to obtain a mapping onto the 2D
classical bond percolation transition, from which we obtain three independent
critical exponents, and universal ratios, exactly.  An anisotropic version of
the model is also mapped onto an antiferromagnetic sl$(2|1)$ SUSY \cite{snr}
quantum spin chain.  The results are in very good agreement with recent
numerical simulations \cite{khac,brad}.

We study the spin QH transition in an alternative description that is obtained
from the superconductor after a particle-hole transformation on the down-spin
particles \cite{sfbn}, which interchanges the roles of particle number and
$z$-component of spin, and so particle number is conserved rather than spin.
This makes it possible to use a single-particle description, at the cost of
obscuring the SU(2) symmetry.  The single-particle energy ($E$) spectrum has a
particle-hole symmetry \cite{az} under which $E\rightarrow -E$, so when states
are filled up to $E=0$, the positive-energy particle and hole excitations become
doublets of the global SU(2) symmetry.  In this picture, a (nonrandom) Zeeman
term $H$ for the quasiparticles maps onto a simple shift in the Fermi energy
to $E=H$ \cite{brad}, splitting the degeneracy.

\begin{figure}
\epsfxsize=3.375in
\centerline{\epsffile{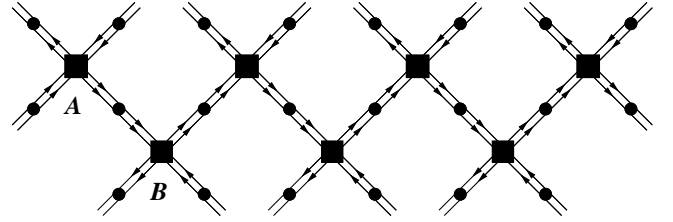}}
\vspace{0.1in}
\caption{The network model.}
\label{network}
\end{figure}

The model \cite{khac,brad} is a network (generalizing Ref.~\cite{cc}), in which
a particle of either spin and with $E=0$, represented by a doublet of complex
fluxes, can propagate in one direction along each link (Fig.  \ref{network}).
The propagation on each link is described by a random SU(2) scattering matrix
(the black dot), with a uniform distribution over the SU(2) group; the absence
of an additional random U(1) phase here is crucial and implies that the global
SU(2) spin-rotation (or particle-hole) symmetry is unbroken.  As in Ref.\
\cite{cc}, there are two sublattices, A and B, on which the nodes are related 
by a 90$^\circ$ rotation.  Scattering of the fluxes at the nodes (black squares) 
is described by orthogonal matrices diagonal in spin indices:  ${\cal S}_S = 
{\cal S}_{S\up} \oplus {\cal S}_{S\down}$,
\beq
{\cal S}_{S\sigma} = \left( \begin{array}{cc} (1-t_{S\sigma}^2)^{1/2} & 
t_{S\sigma}^{\vphantom{2}} \\
-t_{S\sigma}^{\vphantom{2}} &  (1-t_{S\sigma}^2)^{1/2} \end{array} \right),
\eeq
where $S=A$, $B$ labels the sublattice and $\sigma = \up$, $\down$ labels the 
spin direction. The network is spatially isotropic when the 
scattering amplitudes on the two sublattices are related by
$
t_{A\sigma}^2 + t_{B\sigma}^2 = 1.
$
                                                                      
The network has a multicritical point at $t_{A\sigma} = t_{B\sigma} = 2^{-1/2}$
(for the isotropic case).  Taking $t_{A\sigma}\neq t_{B\sigma}$ (but keeping
$t_{S\up} = t_{S\down}$) drives the system through a QH transition between an
insulator and a QH state, and the Hall conductance (now for charge) jumps from
zero to 2.  Making $t_{S\up} \ne t_{S\down}$ breaks the global SU(2) symmetry,
and splits the transition into two ordinary QH transitions \cite{khac} each in
the unitary class.  As we will argue later, this perturbation is different from
the uniform Zeeman term.

We briefly describe, for the present case, the main steps of the SUSY method for
the network models \cite{grs}.  Transport and other properties of the network,
such as its conductance, may be expressed in terms of sums over paths on the
network.  Such a sum may be written in second-quantized language as a
correlation function, $\langle\ldots\rangle\equiv{\rm STr}\,({\cal T}\ldots U)$
where the supertrace contains an evolution operator $U$ of an associated quantum
1D problem, ${\cal T}$ is a time-ordering symbol, and $\ldots$ stands for
operators that represent the ends of paths and correspond physically to density,
current, etc.  In this form, the average can be taken to obtain moments of
physical quantities, and we leave this implicit in later notation.  In this 1D
problem vertical rows of links of the network become sites, and the vertical
direction becomes (imaginary) time (we assume for the present periodic boundary
conditions in both directions).  The evolution operator $U$, composed of
transfer matrices for links and nodes, acts in a tensor product of Fock spaces
of bosons and fermions on each site.  The presence of a fermion or boson on a
link---i.e.  on a site at an instant of discrete time---represents an element of
a path traversing that link \cite{grs}.  Both bosons and fermions are needed to
ensure the cancellation of contributions from closed loops.  Usually one needs
two types of bosons and fermions, retarded and advanced, to be able to obtain
two-particle properties.  However, the particle-hole symmetry relates retarded
and advanced Green's functions \cite{sf}.  Hence, for the study of {\em mean}
values of simple observables, we need only one fermion and one boson per spin
direction per site.  (To study fluctuations and other observables, $N$ types of
fermion and boson are needed, and the SUSY below becomes osp($2N|2N$)
\cite{az}.)  We denote them by $f_\sigma$, $b_\sigma$ for the sites related to
the links going up (up-sites), and $\fb_\sigma$, $\bb_\sigma$ for the
down-sites.  On the up-sites, $f_\sigma$, $b_\sigma$ are canonical, but to
ensure the cancellation of closed loops we must either take the fermions on the
down-sites to satisfy $\{\fbph_\sigma,\fbd_{\sigma'}\} =
-\delta_{\sigma\sigma'}$, or similarly for the bosons.

To begin, we consider the spin-rotation invariant case with $t_{S\up} =
t_{S\down} = t_S$.  In this case, for any realization of the disorder in the
scattering matrices, the transfer matrices commute with the sum over sites of
the eight generators (superspin operators) of the superalgebra sl$(2|1)\cong$
osp$(2|2)$, similarly to Ref.\ \cite{grs}.  The generators for each site are
constructed as all bilinears in the fermions and bosons and their adjoints,
which are singlets under the random SU(2).  These are denoted by \cite{snr} $B$,
$Q_3$, $Q_{\pm} $, $V_{\pm} $, $W_{\pm}$, and have similar expressions for the
two types of sites.  Cancellation of closed loops would only require invariance
under the gl$(1|1)$ subalgebra generated by $B$, $Q_3$, $V_-$, and $W_+$.  The
larger SUSY that exists when $t_{S\up} = t_{S\down}$ is a manifestation of the
global SU(2) symmetry.  

The transfer matrix describing the evolution on a link, after averaging over
the random SU(2) matrices, projects the states on the corresponding site to a
three-dimensional subspace of singlets of the random SU(2)\cite{sf}.  On the
up-sites these form the fundamental representation {\bf 3} of sl$(2|1)$, and we
denote them as $|m\ra$, $m=0$, $1$, $2$.  Similarly, on the down-sites the
three singlet states form the representation $\bbox{\bar 3}$, dual to {\bf 3},
and we call them $|{\bar m}\ra$; $m$ is the number of fermions on a site of
either type.  We find that $|{\bar 1}\ra$ has negative squared norm, $\la {\bar
1}|{\bar 1}\ra = -1$, while the others are positive.  Thus, after averaging, we
have a horizontal chain of sites with alternating dual representations on the
two sublattices and a discrete-time evolution along the vertical direction
given by the transfer matrices at the nodes, which will be specified below.

\begin{figure}
\epsfxsize=3.375in
\centerline{\epsffile{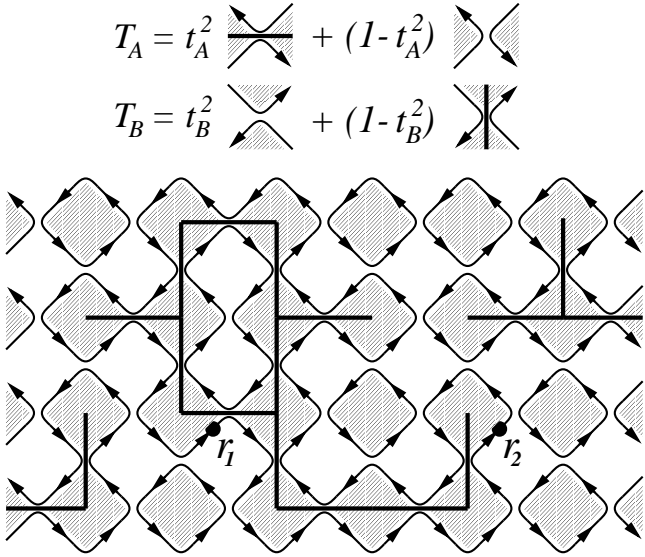}}
\vspace{0.1in}
\caption{Graphical representation of the transfer matrices $T_S$, percolating 
clusters, and two points $r_1$, $r_2$, on a hull.}
\label{cluster}
\end{figure}

We now consider in detail the node transfer matrix $T_S$ on a single node on 
sublattice $S$. After the averaging, it acts in the tensor product 
$\bbox{3} \otimes \bbox{\bar 3}$ for the two sites.
Because of the sl$(2|1)$ SUSY, we find that it takes the form 
\beq
T_S = t_S^2 P_{\bbox{1}} + (1-t_S^2) I\otimes {\bar I}. 
\eeq
Here the first term contains the projection operator $P_{\bbox{1}}=|s\ra\la s|$
onto the normalized singlet state 
$
|s\ra = \sum_m |m\ra \otimes |{\bar m}\ra,
$
while in the second term
$I$, $\bar{I}$ are the identity operators on the two sites (note that 
$
{\bar I} = |{\bar 0}\ra \la {\bar 0}| - |{\bar 1}\ra \la {\bar 1}| +
|{\bar 2}\ra \la {\bar 2}|
$).
The two terms in $T_S$ represent the two ways to 
sl$(2|1)$-invariantly couple the in- and out-going states at the node,
such that the incoming state (in the fundamental representation $\bf 3$)  
flows out unchanged, turning either to the right or the left. They can be 
represented graphically as shown at the top in Fig.
\ref{cluster}. 

When we multiply the transfer matrices together and take the supertrace in the
tensor product of all sites to calculate the partition function $Z={\rm
STr}\,U$, the result is given by the sum of all contributions of closed loops
that fill the links of the network, weighted by factors of either $t_S^2$ or
$(1-t_S^2)$ for each node.  Each loop contributes a factor coming from the sum
over the three states that can propagate around the loop, the supertrace ${\rm
str}\, 1=1$ taken in the fundamental $\bf 3$.  It is also clear that $Z$ is
equal to 1, as it is also before averaging.

The sum over loops on the links of the network is equivalent to the bond
percolation problem on the square lattice, as follows.  In Fig.\ \ref{cluster},
we shade one-half of the plaquettes of the network in checkerboard fashion.
The two terms in $T_S$ possible at each node either do or do not connect the
shaded plaquettes, as indicated by the thick undirected line segments.  At each
$A$- (respectively, $B$-) node we have a horizontal (vertical) line with
probability $p_A = t_A^2$ ($p_B=1-t_B^2$).  Then on the square lattice formed
by the shaded plaquettes we have the classical bond percolation problem, and
the loops are the boundaries (or ``hulls'') of the percolation clusters.  This
SUSY representation of percolation easily generalizes to sl$(n+1|n)$ SUSY,
$n\geq 1$, using the $2n+1$-dimensional fundamental representation and its
dual.

Many critical exponents for 2D percolation are known exactly. First, there is
the correlation length exponent, which immediately gives the localization 
length for the spin QH transition,
\beq
\xi \sim |p_S-p_{Sc}|^{-\nu},
\eeq 
with $\nu=4/3$ \cite{dennijs}; the critical values are $p_{Ac}=p_{Bc}=1/2$ in
the isotropic case. Then, because the basic operators of our system
are the superspins, which act on the states that live on the hulls of the
percolation clusters, we should consider the exponents associated with these
hulls. These include an infinite set of scaling dimensions for the 
so-called $n$-hull operators \cite{sd},
\beq
\label{dim}
x_n = (4n^2 - 1)/12.
\eeq
The exponent $x_n$ describes the spatial decay at criticality 
$\sim |r_1-r_2 |^{-2x_n}$ of the probability that $n$ distinct 
hulls each pass close to each of the two points $r_1$ and $r_2$.
There is also a set of analogous exponents for the same correlators near a
boundary \cite{sb},
\beq
\label{bdim}
{\tilde x}_n = n(2n-1)/3.
\eeq 

We will now relate further physical quantities within our model to percolation
exponents, through the SUSY mapping. We write the superspins as a 
single 8-component object $J$ for either up- or down-sites. These
can be inserted in any links of the network, to obtain a correlator such as 
$\langle Q_3(r_1)Q_3(r_2)\rangle$, where $r_1$ and $r_2$ represent links of 
the network. Then using the same graphical expansion, we obtain a sum over 
loop configurations, now with the positions of the insertions marked on the 
loops, and for loops with insertions, the factor 1 is replaced by a supertrace 
(in the fundamental) of the product of matrices that represent the $J$'s 
inserted. We then require only the total probabilities that loops pass through 
the marked points in various ways. The simplest example is a two-point 
function of $J$'s, which is nonvanishing only if the $J$'s are on the same 
loop, because ${\rm str}\, J=0$ for all components of $J$. The leading term in 
the probability that the two points are on the same loop (hull) is governed by 
the leading $1$-hull operator in the continuum theory \cite{sd}, giving
\beq
\langle (-1)^{i_1+i_2}J(r_1)J(r_2)\rangle\sim |r_1-r_2|^{-2x_1},
\eeq
at the transition, where $x_1=1/4$ as specified above. The reason for 
the staggering factors $(-1)^{i_1}$, where $i_1$ is the site corresponding to 
$r_1$, will become clear momentarily.

It is useful to consider the anisotropic limit of the model.  This is defined by
$t_A$, $t_B\rightarrow 0$ with a fixed ratio $t_A/t_B=\epsilon$.  Then the
transfer matrices $T_S$ may be expanded in $t_S$ and recombined in the
exponential.  The evolution operator has the form $U\simeq \exp(-2 t_A t_B \int
\!\!  d\tau {\cal H}_{\text{1D}})$, where the effective Hamiltonian ${\cal 
H}_{\rm 1D}$ describes a 1D superspin chain, with alternating $\bf 3$ and 
$\bf{\bar{3}}$ representations, and continuous imaginary time $\tau$:
\beq
{\cal H}_{\rm 1D} = \sum\nolimits_i \left( \epsilon \, J_{2i-1} \cdot J_{2i} + 
\epsilon^{-1} J_{2i} \cdot J_{2i+1} \right).
\eeq
Here $J\cdot J$ denotes the sl$(2|1)$ invariant product \cite{snr}.  The
transition point, where ${\cal H}_{\rm 1D}$ for an infinitely-long chain is 
gapless, is now at $\epsilon=1$.  The two-site version of ${\cal H}_{\rm 1D}$ 
appeared in  Ref.\ \cite{sf}.

The sum $\sum_i J_i$ is the generator of global SUSY transformations, and so
$J_i$, viewed as a function of $i$, is the superspin density on the lattice,
which gives a subleading contribution $\sim r^{-2}$ to the $J$-$J$ correlation
at criticality.  The 1-hull operator must therefore be the staggered part,
$(-1)^i J_i$.

The 1-hull operators represented by $(-1)^iJ_i$ have several physical 
applications. Components such as $Q_+=f_\up^\dagger f_\down^\dagger$ on the 
up-sites create fermions, so produce ends for the quasiparticle paths. 
The sum of all such paths between $r_1$ and $r_2$ represents the quasiparticle 
Green's function, $G$. To obtain nonzero results on averaging, we must 
multiply the retarded and advanced Green's functions before averaging, but 
this can be replaced by a spin-singlet combination of our fermions or bosons 
\cite{sf}. The staggered part of this averaged correlator represents the 
average zero-frequency density-density (``diffusion'') propagator 
$\overline{|G|^2}$ (and also the average conductance between two point-contacts 
\cite{jmz}), which therefore falls as $|r_1-r_2|^{-1/2}$ at the transition. 
Moreover, the local density of states $\rho(r,E)$ is represented by another 
component of the 1-hull operator, because both it and the density operator 
contain wavefunctions squared, $\sim|\psi|^2$, in the original problem and 
so scale in the same way. The energy $E$ itself (set to zero hitherto) has 
scaling dimension $y_1=2-x_1$ because an imaginary part $\eta$ of $E$ induces a 
staggered ``magnetic field'' term $\sum_i\eta\rho_i(i\eta)$ in ${\cal H}_{\rm 
1D}$. Hence for the average we have at criticality
\beq
\overline{\rho(r,E)} \sim |E|^{x_1/y_1} = |E|^{1/7}.
\eeq
Also, since a uniform Zeeman term $H$ causes a shift in the Fermi energy, it
induces a correlation length $\xi_H\sim |H|^{-\nu_1}$, where $\nu_1=1/y_1=4/7$. 

We have already identified the value $\nu=4/3$ of the localization length
exponent $\nu$ with that in percolation. In terms of ${\cal H}_{\rm 1D}$, the 
effect of a small deviation  $\delta \equiv \epsilon -1$ is to add the 
perturbation
$
\delta \sum_i(-1)^iJ_i\cdot J_{i+1}
$
to the critical ${\cal H}_{\rm 1D}$.  This term contains the dimer operator
$D_i=(-1)^iJ_i\cdot J_{i+1}$, which is odd under reflection through any lattice
site (parity).  The scaling dimension $x_2$ of the 2-hull operator is the same
as that of this ``thermal'' perturbation for the transition, that is
$\nu=\nu_2=1/y_2$, $y_2=2-x_2$ \cite{sd}.  We therefore expect that the 2-hull
operator is part of a multiplet of staggered two-superspin operators, that are
similar to $D_i$, but are not all sl$(2|1)$ singlets.

As a final perturbation of the critical Hamiltonian, we consider the effect of
$t_{S\up}\neq t_{S\down}$. This breaks the global SU(2) symmetry, and 
breaks the SUSY to gl$(1|1)$. Taking $t_{A\sigma}=t_{B\sigma}$, 
we find that the effect is to add to ${\cal H}_{\rm 1D}$ a term
$
(t_\up-t_\down)^2\sum_i\hat{J}_i\cdot\hat{J}_{i+1},
$
where $\hat{J}_i$ is the 4-component set of generators of gl$(1|1)$, and
the product is invariant under this algebra. This term is an anisotropy 
in superspin space. The two QH transitions it produces \cite{khac} cannot be 
seen in our formulation without explicitly introducing both retarded and 
advanced fermions and bosons, and we will see only exponentially decaying 
correlations. The correlation length $\xi_\Delta$ induced by 
$\Delta=t_\up-t_\down$ scales as
\beq
\xi_\Delta \sim |\Delta|^{-\mu},
\eeq  
in the notation of Ref.\ \cite{khac}, for small $\Delta$.  If the spin
anisotropy $\hat{J}_i\cdot\hat{J}_{i+1}$ has dimension $x'$, then we will have
$\mu=2/(2-x')$.  The operator does not appear to be the 1-hull operator, and has
the opposite parity to the 2-hull.  However, the operator product of two 1-hull
operators has the correct parity and might contain this operator.  In conformal
field theory, the 1-hull operator can be represented by $\phi_{2,2}$ in the Kac
classification of $c=0$ Virasoro representations.  The fusion rules for this
primary field with itself contain the leading nontrivial operator $\phi_{1,3}$,
which we view as a subleading 1-hull operator, with scaling dimension
$\hat{x}_1=2h_{1,3}=2/3$.  We suggest that $x'=\hat{x}_1=2/3$, which yields
$\mu=3/2$.  We further suggest that this operator describes a {\it random}
Zeeman term (with zero mean).

Finally, we note that the average two-probe conduc\-tance of our system with 
open ends \cite{grs}, and with $t_{S\up}=t_{S\down}$, can be related to the 
number $n$ of hulls that connect one end to the other (and back). Each such 
configuration of loops contributes $n$ to the conductance, times $2$ for spin, 
so the mean conductance has the scaling form
\beq
\bar{g} = 2 \sum_{n=1}^\infty n P(n,L/W, L/\xi),
\eeq 
where $P(n,L/W,L/\xi)$ is the probability that exactly $n$ hulls run from 
one end to the other and back, for a system of size $L$ by $W$. This can be
considered both for periodic and reflecting transverse boundary conditions. 
At the transition, $\xi=\infty$, and for large $L/W$, it is known \cite{cardy} 
that $P(n,L/W,0)\sim e^{-2\pi x_n L/W}$ for periodic, and $\sim e^{-\pi 
\tilde{x}_n L/W}$ for reflecting boundaries. The sum for $\bar{g}$ is 
dominated by the $n=1$ term in this limit, so it has the form $\bar{g}\sim 
e^{-L/\xi_{\rm 1D}}$, giving the behavior of the localization length 
$\xi_{\rm 1D}$, the only parameter that enters in the complete distribution of 
conductance in this limit \cite{b}. As $L/W\rightarrow 0$, we expect that 
$\bar{g}\propto W/L$, implying that there is a nonzero critical conductivity.

We may now compare our results with those of recent numerical work.  In Ref.\
\cite{khac}, the results obtained were $\nu\simeq 1.12$ and $\mu\simeq 1.45$.
These are in fair agreement with our predictions, especially for $\mu$ where
our theoretical argument is less well established.  The authors of Ref.\
\cite{brad} study the SUSY spin chain numerically, and find critical exponents
$x_1 = 0.26 \pm 0.02$ and $x_2 = 1.24 \pm 0.01$, in excellent agreement with our 
predictions.

To conclude, we have used SUSY methods to find a remarkable equivalence of a
quasiparticle localization problem, the spin quantum Hall transition, to 2D
percolation, resulting in the exact values of three exponents, and the
universal ratios for the localization length in the 1D limit.

We are grateful to T.  Senthil, J.  B.  Marston, and M.  P.  A.  Fisher for
useful discussions, and for sharing numerical results before publication.  We
also thank J.  Chalker, V.  Gurarie and M.  Zirnbauer for helpful discussions.
This work was supported by the NSF under grants No.  PHY94-07194, DMR-9528578
(IAG), DMR-9157484 (NR), and in part by the A.  P.  Sloan Foundation (AWWL).

\end{document}